\documentclass[sigconf]{acmart}
\usepackage{booktabs} %

\setcopyright{rightsretained}
\acmConference[PASC'19]{Platform for Advanced Scientific Computing}{June 2019}{ETH Zurich, Switzerland}
\copyrightyear{2019}
\usepackage[ruled, linesnumbered]{algorithm2e}
\usepackage{algorithmic}

\SetCommentSty{mycommfont}

\usepackage{framed} 

\immediate\write18{%
  makeindex -s nomencl.ist -o \jobname.nls -t \jobname.nlg \jobname.nlo%
}

\usepackage{multicol}
\usepackage{multirow}

\usepackage{listings}
\usepackage{color}
\definecolor{dkgreen}{rgb}{0,0.6,0}
\definecolor{gray}{rgb}{0.5,0.5,0.5}
\definecolor{mauve}{rgb}{0.58,0,0.82}
\definecolor{darkmagenta}{rgb}{0.98,0.15,0.45}
\definecolor{saffron}{rgb}{1,0.6,0.2}
\usepackage{tikz}
\usetikzlibrary{patterns}
\usepackage{pgfplots}
\pgfplotsset{compat=1.14}
\usepgfplotslibrary{patchplots}
\tikzstyle{class} = [rectangle, minimum width=1cm, minimum height=0.1cm, text centered, fill=white!10]
\tikzstyle{arrow} = [thick,->,>=stealth]
\tikzstyle{darrow} = [dashed,->,>=stealth]

\usetikzlibrary{decorations.shapes}
\tikzset{decorate sep/.style 2 args={decorate, decoration={shape backgrounds, shape=circle, shape size=#1, shape sep=#2} }}

\usepackage{lipsum}
\usepackage{amsmath}
\usepackage{subcaption}
\usepackage{amsbsy}

\newcommand{\bsigma}{{\boldsymbol{\sigma}}}
\newcommand{\rd}{\mathrm{d}}
\newcommand{\bc}{\mathbf{c}}
\newcommand{\bx}{\mathbf{x}}

\newcommand{\bu}{\mathbf{u}}

\newcommand{\fft}{{\texttt{fft}}}
\newcommand{\ifft}{{\texttt{ifft}}}

\newcommand{\Is}{{(p)}}
\newcommand{\Js}{{(q)}}
\newcommand{\IJs}{{(pq)}}
\newcommand{\is}{p}
\newcommand{\js}{q}
\newcommand{\mi}{m_\is}
\newcommand{\mj}{m_\js}

\newcommand{\ijs}{{pq}}
\newcommand{\jis}{{qp}}
\newcommand{\one}{{(1)}}
\newcommand{\two}{{(2)}}

\newcommand{\Kn}{\mathrm{Kn}}
\newcommand{\pre}{\mathrm{(pre)}}
\newcommand{\q}{\rho}

\begin{document}
\title{A discontinuous Galerkin fast spectral method for multi-species full Boltzmann on streaming multi-processors}

\subtitle{Accepted}
\subtitlenote{Accepted on 20th Feb 2019, To be presented at PASC between 12-14th June 2019}

\author{Shashank Jaiswal}
\affiliation{%
  \institution{Purdue University}
  \streetaddress{701 W. Stadium Avenue}
  \city{West Lafayette}
  \state{Indiana}
  \postcode{47907}
  \country{U.S.A}
}
\email{jaiswal0@purdue.edu}

\author{Jingwei Hu}
\affiliation{%
  \institution{Purdue University}
  \streetaddress{150 N. University St}
  \city{West Lafayette}
  \state{Indiana}
  \postcode{47907}
  \country{U.S.A}
}
\email{jingweihu@purdue.edu}

\author{Julien K. Brillon}
\affiliation{%
  \institution{Purdue University}
  \streetaddress{701 W. Stadium Avenue}
  \city{West Lafayette}
  \state{Indiana}
  \postcode{47907}
  \country{U.S.A}
}
\email{jbrillon@purdue.edu}

\author{Alina A. Alexeenko}
\affiliation{%
  \institution{Purdue University}
  \streetaddress{701 W. Stadium Avenue}
  \city{West Lafayette}
  \state{Indiana}
  \postcode{47907}
  \country{U.S.A}
}
\email{alexeenk@purdue.edu}
\renewcommand{\shorttitle}{}

\begin{abstract}
When the molecules of a gaseous system are far apart, say in microscale gas flows where the surface to volume ratio is high and hence the surface forces dominant, the molecule-surface interactions lead to the formation of a local thermodynamically non-equilibrium region extending few mean free paths from the surface. The dynamics of such systems is accurately described by Boltzmann equation. However, the multi-dimensional nature of Boltzmann equation presents a huge computational challenge. With the recent mathematical developments and the advent of petascale, the dynamics of full Boltzmann equation is now tractable. We present an implementation of the recently introduced multi-species discontinuous Galerkin fast spectral (DGFS) method for solving full Boltzmann on streaming multi-processors. The present implementation solves the inhomogeneous Boltzmann equation in span of few minutes, making it at least two order-of-magnitude faster than the present state-of-art stochastic method---direct simulation Monte Carlo---widely used for solving Boltzmann equation. Various performance metrics, such as weak/strong scaling have been presented. A parallel efficiency of 0.96--0.99 is demonstrated on 36 Nvidia Tesla-P100 GPUs.
\end{abstract}



\maketitle

\section{Introduction}
From the fundamental mass/momentum conservation principles, it can be inferred that, in the presence of external forces, say, pressure and temperature gradients, the heavier species moves slower and the lighter species moves faster giving rise to a phenomena termed as diffusion, and effects thereof. Diffusion processes are critical in many applications, for instance, the measurement of the neutrino mass using a windowless gaseous tritium source in the ongoing KATRIN experiment \cite{bornschein2005katrin}. The dynamics of such systems (and others) are governed by the Boltzmann equation---an integro-differential equation describing the evolution of the distribution function in six-dimensional phase space---which models the dilute gas behavior at the molecular level to accurately describe a wide range of non-continuum flow phenomena, for instance, shocks, expansions into vacuum \cite{Muntz1989} as well as velocity and thermal slip at gas-solid interfaces \cite{sharipov2003velocity,sharipov2004velocity}. Most rarefied flows of technological interest involve gas mixtures with species diffusion playing a decisive role in turbulent, chemically reacting flows, and evaporation/condensation processes \cite{takata2007half}. 

The approaches for numerical solution of the Boltzmann equation date back to as early as 1940s \cite{grad1949} using, for example, the now widely used direct simulation Monte Carlo (DSMC) method \cite{Bird}. The DSMC method, based on the kinetic theory of dilute gases, models the binary interactions between particles stochastically. However, it is this stochastic nature, that makes the method unsuitable for flows involving species in trace concentration, for instance, to analyze the spectrum of beta electrons emitted by tritium source which can be substantially different in the presence of the impurities in KATRIN experiment \cite{bornschein2005katrin,sharipov2005separation}. 

The main difficulty of numerically solving the full Boltzmann equation lies in its complicated collision term. Recently, a fast Fourier spectral method for the multi-species Boltzmann collision operator was introduced in \cite{jaiswal2019dgfsMulti}. The complexity for a single evaluation of the collision operator is reduced from $O(N^6)$ (direct calculation) to $O(M N_\rho N^3 \log N)$ (based on a low-rank decomposition strategy), where $N$ is the number of discretization points in each velocity dimension, $N_\rho \sim O(N)$ is the number of discretization points in the radial direction needed for low-rank decomposition, and $M \ll N^2$ is the number of discretization points on the sphere. Based on \cite{JAH19}, a discontinuous Galerkin fast spectral (DGFS) method was also proposed in \cite{jaiswal2019dgfsMulti} for solving the full multi-species Boltzmann equation. DGFS can produce high order spatially and temporally accurate solutions for low-speed and unsteady flows in micro-systems, and is amenable to excellent nearly-linear scaling characteristics on massively parallel architectures. This paper focuses on implementation aspects of multi-species DGFS with an emphasis on establishing the algorithmic behavior of such numerical schemes. More specifically, here, we are concerned about the scaling characteristics of DGFS on multi-GPU/multi-CPU systems.

In the section that follows, we describe the multi-species Boltzmann in brief, followed by a description of the collision operator algorithm. Various performance metrics such as weak/strong scaling, and micro benchmarks involving, static adaptivity of cell-size and polynomial order approximation for rarefied gas-flows have been discussed in section \ref{section_benchmarks}. Concluding remarks are given in section~\ref{section_conclusions}. 

\section{The multi-species Boltzmann equation}
The \textit{non-dimensional} Boltzmann equation for multi-species, mono-atomic gas without external forces can be written as (cf. \cite{jaiswal2019dgfsMulti}) 
\begin{align} \label{eqn_bzeNonDim}
\frac{\partial f^\Is}{\partial t}+\bc\cdot \nabla_{\hat{\bx}} f^\Is= \sum_\js \frac{1}{\mathrm{Kn}_{\ijs}}\mathcal{Q}^\IJs
, \quad \is=1,2,\dots,n,
\end{align}
where $n$ denotes number of species in the mixture -- each of them represented by a number distribution function $f^\Is(t,\bx,\bc)$ of time $t$, position $\bx$, and particle velocity $\bc$. The collision operator $\mathcal{Q}^\IJs$ takes into account interactions between species $\is$ and $\js$, which acts only in the velocity space:
\begin{align} 
\mathcal{Q}^\IJs(f^\Is,f^\Js)(\bc) &=  \int_{\mathbb{R}^3}\int_{S^{2}}B_{\ijs}(|\bc-\bc_*|,\bsigma\cdot \widehat{(\bc-\bc_*)}) \nonumber \\
&\left[f^\Is(\bc')f^\Js(\bc_*')-f^\Is(\bc)f^\Js(\bc_*)\right]\rd{\bsigma}\,\rd{\bc_*},
\label{eqn_bzeDim}
\end{align}
where $(\bc, \bc_*)$ and $(\bc', \bc'_*)$ denote the pre and post collision velocity pairs, which are related through momentum and energy conservation as
\begin{align}
\left\{
\begin{array}{l}
\displaystyle \bc'=\frac{\bc+\bc_*}{2}+\frac{(1-\mj/\mi)}{2(1+\mj/\mi)}(\bc-\bc_*)+\frac{1}{(\mi/\mj+1)}|\bc-\bc_*|\bsigma, \\[12pt]
\displaystyle \bc_*'=\frac{\bc+\bc_*}{2}+\frac{(1-\mj/\mi)}{2(1+\mj/\mi)}(\bc-\bc_*)-\frac{1}{(\mj/\mi+1)}|\bc-\bc_*|\bsigma,
\end{array}\right.
\end{align}
where $m_\is$, $m_\js$ denote the mass of particles of species $\is$ and $\js$ respectively. Here, the vector $\bsigma$ varies over the unit sphere $\mathcal{S}^2$. The quantity $\mathcal{B}_\ijs$ ($\geq 0$) is the collision kernel depending only on $| \bc - \bc_*|$ and the scattering angle $\chi$ (angle between $\bc-\bc_*$ and $\bc'-\bc'_*$). In the present work, we consider the variable soft sphere (VSS) \cite{koura1991variable} scattering model. It is worth emphasizing that although the VSS collision kernel is adopted in the present work for easy comparison with DSMC solutions, the fast spectral method we use for the collision operator applies straightforwardly to general collision kernels (see \cite{GHHH17,JAH19,jaiswal2019dgfsMulti}).

For Variable Soft-Sphere model \cite{Bird} in particular, the non-dimensional collision-kernel $B_\IJs$, and the Knudsen number $\mathrm{Kn}_\ijs$  are given as
\begin{align} \label{eq_VSSNonDim}
B_{\IJs} &= \frac{1}{\sqrt{1+\mi/\mj}} \frac{1}{\Bigl(\frac{\mi \mj}{\mi + \mj}\Bigr)^{(\omega_{\ijs}-0.5)}} \nonumber \\ &\frac{\alpha_{\ijs}}{2^{1+\alpha_{\ijs}}\,\Gamma(2.5-\omega_\ijs)\pi}|\bc - \bc_*|^{2(1 - \omega_{\ijs})} \; (1 + \cos \chi)^{\alpha_{\ijs}-1},
\end{align}
\begin{align} \label{eq_Kn}
\mathrm{Kn}_{\ijs} &= \frac{1}{\sqrt{1 + \mi/\mj}\; \pi\; n_0\; d^2_{(\mathrm{ref}, \ijs)}\; (T_{\text{ref},\ijs}/T_0)^{\omega_{\ijs}-0.5}\; H_0}.
\end{align}
Here $\Gamma$ denotes the usual Gamma function, $d_{(\mathrm{ref},\ijs)}$, $T_{(\mathrm{ref},\ijs)}$, $\omega_\ijs$, and $\alpha_\ijs$ are, respectively, the reference diameter, the reference temperature, the viscosity index, and the scattering parameter. The diameter $d_{(\mathrm{ref},\ijs)}$ and exponent $\alpha_\ijs$ are determined so that the transport (viscosity and diffusion) coefficients of VSS are consistent with experimental data. Additionally $H_0$, $T_0$, $n_0$, and $m_0$, respectively, denote the characteristic length,  characteristic temperature, characteristic number density, and characteristic mass $m_0$. Based upon these, we define the characteristic velocity as $u_0 = \sqrt{2 k_B T_0/m_0}$ where $k_B$ refers to Boltzmann constant; and characteristic time as $t_0=H_0/u_0$. For convenience, we define a pre-factor $\beta^\IJs$ as
\begin{align}
\beta^{\IJs} &= \frac{1}{\Kn_\ijs} \frac{1}{\sqrt{1+\mi/\mj}} \frac{1}{\Bigl(\frac{\mi \mj}{\mi + \mj}\Bigr)^{(\omega_{\ijs}-0.5)}} \frac{\alpha_{\ijs}}{2^{1+\alpha_{\ijs}}\,\Gamma(2.5-\omega_\ijs)\pi}
\end{align}
Henceforth, we will always refer to the non-dimensional Boltzmann equation (\ref{eqn_bzeNonDim}) in our presentation.

\subsection{The collision operator}
First, note that $\mathcal{Q}^\IJs(f^\Is,f^\Js)$ does not depend on spatial coordinate $\mathbf{x}$. Given distribution functions $f^\Is$ and $f^\Js$ of species $\is$ and $\js$, dependent only on the velocity coordinate $\bc$: discretized \textit{uniformly} using $N^3$ points, the method produces $\mathcal{Q}^\IJs(f^\Is,f^\Js)$ at the same grid with $O(M N_\rho N^3\log N)$ complexity, where $N_\rho \sim O(N)$ is the number of Gauss-Legendre quadrature/discretization points in the radial direction needed for low-rank decomposition , $M \ll N^2$ is the number of discretization points on the sphere. The steps (based on \cite{jaiswal2019dgfsMulti}) for evaluating $\mathcal{Q}^\IJs$ can be summarized as:
\begin{itemize}
\item Change the variable $\bc_*$ to $\bu = \bc-\bc_*$:
\begin{align}
    \mathcal{Q}^\IJs(f^\Is,f^\Js)(\bc) &= \int_{\mathbb{R}^3} \int_{\mathcal{S}^2} B_\ijs ( |\bu|, \bsigma\cdot \hat{\bu}) \nonumber \\
    &\Bigl[  f^\Is( \bc')  f^\Js( \bc'_*) -  f^\Is( \bc) f^\Js(\bc-\bu) \Bigr] \rd{\bsigma} \rd{\bu},
\label{eq_apndx_Q}
\end{align}
where $\hat{\bu}$ is the unit vector along $\bu$, and
\begin{align}
\left\{
\begin{array}{l}
\displaystyle \bc'=\bc-\frac{\mj}{\mi+\mj}\bu+\frac{\mj}{\mi+\mj}|\bu|\bsigma, \\[12pt]
\displaystyle \bc_*'=\bc-\frac{\mj}{\mi+\mj}\bu-\frac{\mi}{\mi+\mj}|\bu|\bsigma.
\end{array}\right.
\end{align}
\item Determine the extent of velocity domain $D_L=[-L,L]^3$, and periodically extend $f$, $g$ to $\mathbb{R}^3$.
\item Truncate the integral in $\bu$ to a ball $B_R$ with 
\begin{align}
\displaystyle R=\frac{4}{1+\max(4\mj/(\mi+\mj),2)+\sqrt{1+\mj/\mi}}L
\end{align}
\item Approximate $f^\Is$, $f^\Js$ by truncated Fourier series
\begin{equation}
f^\Is(\bc)=\sum_{k=-N/2}^{N/2-1}\hat{f}^\Is_k e^{i\frac{\pi}{L}{k}\cdot \bc}, \;\;
f^\Js(\bc)=\sum_{k=-N/2}^{N/2-1}\hat{f}^\Js_k e^{i\frac{\pi}{L}{k}\cdot \bc}.
\end{equation}
Note here $k$ is a three-dimensional index.
\item Substitute $f^\Is$, $f^\Js$ into (\ref{eq_apndx_Q}), and perform the standard Galerkin projection
\begin{equation}
\begin{split}
\hat{\mathcal{Q}}^\IJs_{k}:&=\frac{1}{(2L)^3}\int_{D_L} \mathcal{Q}^\IJs(f^\Is,f^\Js)(\bc)e^{-i\frac{\pi}{L}k\cdot \bc}\rd{\bc}\\
&=\sum_{\substack{l,m=-N/2\\l+m=k}}^{N/2-1}\Bigl[G^{\IJs +}(l,m)-G^{\IJs -}(m,m)\Bigr]\;\hat{f}^\Is_l\hat{f}^\Js_m,
\end{split}
\label{eq_apndx_Qprojected}
\end{equation}
where $k=-N/2,\dots,N/2-1$, and the kernel modes $G^{\IJs+}$ and $G^{\IJs-}$ are given by
\begin{align}
G^{\IJs+}(l,m)&=\int_{\mathcal{B}_R}\int_{S^{2}}B_\ijs(|\bu|,\bsigma\cdot \hat{\bu}) \nonumber\\ 
&\left[e^{-i\frac{\pi}{L}\frac{\mj}{\mi+\mj}(l+m)\cdot \bu+i\frac{\pi}{L}|\bu|\left(\frac{\mj}{\mi+\mj}l-\frac{\mi}{\mi+\mj}m\right)\cdot \bsigma}\right]\rd{\bsigma}\,\rd{\bu} \nonumber\\
G^{\IJs-}(m,m)&=\int_{\mathcal{B}_R}\int_{S^{2}}B_\ijs(|\bu|,\bsigma\cdot \hat{\bu})\left[e^{-i\frac{\pi}{L}m \cdot \bu}\right]\rd{\bsigma}\,\rd{\bu}.
\label{GG}
\end{align}
\end{itemize}

It is clear that the direct evaluation of $\hat{\mathcal{Q}}^\IJs_k$ (for all $k$) would require $O(N^6)$ complexity. But if we can find a low-rank, separated expansion of $G^{\IJs+}(l,m)$ as
\begin{equation} \label{lowrank}
G^{\IJs+}(l,m) \approx \sum_{r=1}^{N_{\rho}} \alpha_r (l+m) \; \beta_r(l) \; \gamma_r(m),
\end{equation}
then the gain term (positive part) of $\hat{\mathcal{Q}}^\IJs_k$ can be rearranged as
\begin{equation}
\hat{\mathcal{Q}}^{\IJs+}_k = \sum_{r=1}^{N_{\rho}} \alpha_r (k) \sum_{\substack{l,\;m=-N/2 \\ l+m=k}}^{N/2-1} \; \left(\beta_r(l) \hat{f}^\Is_l \right) \; \left(\gamma_r(m) \hat{f}^\Js_m \right),
\label{eq_apndx_fs_Glm_Qk}
\end{equation}
which is a convolution of two functions $\beta_r(l) \hat{f}^\Is_l$ and $\gamma_r(m) \hat{f}^\Js_m$, hence can be computed via fast Fourier transform (FFT) in $O(N_\rho N^3 \log N)$ operations. Note that the loss term (negative part) of $\hat{\mathcal{Q}}^{\IJs-}_k$ is readily a convolution and can be computed via FFT in $O(N^3\log N)$ operations.

In order to find the approximation in (\ref{lowrank}), we simplify (\ref{GG}) as 
\begin{align} \label{realGG}
G^{\IJs+}(l,m) =& \int_{\mathcal{B}_R}\int_{S^{2}}B_\ijs(|\bu|,\bsigma\cdot \hat{\bu}) \nonumber \\ 
&e^{-i\frac{\pi}{L}\frac{\mj}{\mi+\mj}(l+m)\cdot \bu+i\frac{\pi}{L}|\bu|\left(\frac{\mj}{\mi+\mj}l-\frac{\mi}{\mi+\mj}m\right)\cdot \bsigma}\rd{\bsigma}\,\rd{\bu}\nonumber\\
=\int_0^R\int_{S^{d-1}} & F^\IJs(l+m,\rho,\bsigma)e^{i\frac{\pi}{L}\rho\left(\frac{\mj}{\mi+\mj}l-\frac{\mi}{\mi+\mj}m\right)\cdot \bsigma}\rd{\bsigma}\,\rd{\rho},
\end{align}
where
\begin{align}
F^{\IJs}(l+m,\rho,\bsigma)=\rho^{2}\int_{S^{2}}B_\ijs(\rho,\bsigma\cdot \hat{\bu})e^{-i\frac{\pi}{L}\rho\frac{\mj}{\mi+\mj}(l+m)\cdot \hat{\bu}}\,\rd{\hat{\bu}},
\label{eq_apndx_Fkr_vss}
\end{align}
while for the loss term, 
\begin{align}
G^{\IJs-}(m)&=\int_{\mathcal{B}_R}\int_{S^{2}}B_\ijs(|\bu|,\bsigma\cdot \hat{\bu})\;e^{-i\frac{\pi}{L}m \cdot \bu}\;\rd{\bsigma}\,\rd{\bu}\nonumber\\
&=\int_0^R\int_{S^{2}}\int_{S^{2}}\rho^{2} \; B_\ijs(\rho,\bsigma\cdot \hat{\bu})\;e^{-i\frac{\pi}{L}\rho\, m \cdot \hat{\bu}}\;\rd{\bsigma}\,\rd{\hat{\bu}}\,\rd{\rho}.
\label{eq_apndx_Gmm_vss}
\end{align}

For details on the error introduced from the Fourier-spectral approximation, the reader is referred to \cite{jaiswal2019dgfsMulti}. This discussion has been omitted in the present work for brevity.

\subsection{The collision operator algorithm}
The collision operator procedure described above is applicable for general collision kernels for $n$-species mixture. However, for a concise description of the algorithmic ideas from an implementation viewpoint, we restrict our discussion to Variable Soft Sphere collision kernel (\ref{eq_VSSNonDim}). The ideas, however, can certainly be carried over to other collision kernels. 

In multi-species implementation, with the high amount of involved computation, our motive is to avoid spurious computation for every timestep. We first outline the procedure for pre-computing variables that can be stored and reused during the course of the simulation.
\begin{itemize}
\item First, we precompute $(\pi/L\;\rho\;l \cdot \bsigma)$. We use Gauss-Legendre-Quadrature (GLQ) for integration. So $\rho$, the GLQ zeros, is an array of size $N_\q$ (since the integrand oscillates on the scale of O(N), the total number of quadrature points needed should be $\sim O(N)$). Additionally, we use \textit{spherical design} \cite{Womersley} quadrature on sphere. So, $\bsigma$, the spherical-quadrature zeros, is an array of size $M$. $l$ as previously defined is the 3-D velocity-space index, and is therefore an array of size $N^3$. Based upon these $(\pi/L\;\rho\;l \cdot \bsigma)$ is precomputed and stored as a $N_\q \times M \times N^3$ flattened row-major array $\mathrm{a}_{xyz}$. This is described in steps 1--9 of Algo.~(\ref{algo_fastSpectral_jaiswal2018_precompute}).
\item Second, we compute $F(l+m,\rho,\bsigma)$ as per Eq.~(\ref{eq_apndx_Fkr_vss}). Note that $k=l+m$ is velocity-space index of size $N^3$. Since $l+m$, $\rho$, and $\bsigma$ do not change with time, the term $F(l+m,\rho,\bsigma)$ is precomputed and stored as a $N_\q \times M \times N^3$ flattened row-major array $\mathrm{b}^\IJs_{xyz}$ for every collision pair $(\is,\js)$. This is described in step 13 of Algo.~(\ref{algo_fastSpectral_jaiswal2018_precompute}). 
\item Third, we perform precomputation needed for loss-term $G^{\IJs-}(m)$ as per Eq.~(\ref{eq_apndx_Gmm_vss}). The output is stored as a $N^3$ flattened row-major array $\mathrm{c}^\IJs_{z}$ for every collision pair $(\is,\js)$. This is described in step 14 of Algo.~(\ref{algo_fastSpectral_jaiswal2018_precompute}).
\end{itemize}
\begin{algorithm}[!ht]
	\begin{algorithmic}[1]
		\renewcommand{\algorithmicrequire}{\textbf{Input:}}
		\renewcommand{\algorithmicensure}{\textbf{Output:}}
		\REQUIRE Number of points in each-direction of velocity mesh $N$, number of quadrature points for low-rank decomposition $N_\q$, number of points on half-sphere $M$, number of points on pre-computation sphere $M^\pre$, spherical quadrature weight $w_\bsigma$, spherical quadrature-points $\bsigma$ (vector-field size: $M$), pre-computation spherical quadrature weight $w^\pre_\bsigma$, pre-computation spherical quadrature-points $\bsigma^\pre$ (vector-field size: $M^\pre$), Gauss quadrature-weights $w_\rho$ (size: $N_\q$), Gauss quadrature-points $\rho$ (size: $N_\q$), first collision parameter $\gamma_\ijs=2(\omega_\ijs-1)$, second collision parameter $\eta_\ijs=(\alpha_\ijs-1)$, size of velocity mesh $L$, normalized mass $\mi,\mj$ of species-pair $(\is,\js)$
		\ENSURE  a,b,c \\
		\textit{Declare}: \\
			\quad a (size: $MN_\q N^3$), b$^\IJs$ (size: $M N_\q N^3$), c$^\IJs$ (size: $N^3$) \\
			\quad $l$ (vector-field size: $N^3$), v (size: $N$) \\
		\FOR {$x = 0$ to $N-1$}
			\STATE v$_x$ = $x$ - ($x$ $\geq$ N/2) $\times$ N 
		\ENDFOR \\
		\tcp{See octave function: [lx,ly,lz]=ndgrid(v)}
		\STATE $l$ $\gets$ \texttt{ndgrid}(v) \\
		\tcp{Subscript x,y,z on symbols denote array-index}
		\FOR {$x = 1$ to $N_\q$}
		\FOR {$y = 1$ to $M$}
		\FOR {$z = 1$ to $N^3$}
		\STATE a$_{xyz}$ $\gets$ $\pi$/L $\times$ $\rho_x$ $\times$ ($l_z \cdot \bsigma_y$) \\
		\tcp{ ( $\cdot$ ) denotes vector dot-product} 
		\ENDFOR
		\FOR {$\hat{y} = 1$ to $M^\pre$}
		\STATE B$_\ijs$ $\gets$ $\bigl(1 + \bsigma_y \cdot \bsigma^\pre_{\hat{y}}\bigr)^{\eta_\ijs}$ \\
		\FOR {$z = 1$ to $N^3$}
		\STATE b$^\IJs_{xyz}$ $\gets$ b$^\IJs_{xyz}$ + B$_\ijs\;\times\;{w^\pre_\bsigma} \times \rho_x^{\gamma_\ijs+2} \times \exp(\texttt{-1i}\;\times\;\mj/(\mi+\mj)\;\times\;\pi$/L $\times$ $\rho_x$ $\times$ ($l_z \cdot \bsigma^\pre_{\hat{y}})$) \\
        \STATE c$^\IJs_{z}$ $\gets$ c$^\IJs_{z}$ + $({w_{\rho}})_x$ $\times$ $w_\bsigma$ $\times$ B$_\ijs\;\times\;{w^\pre_\bsigma} \times \rho_x^{\gamma_\ijs+2} \times \exp(\texttt{-1i}\;\times\;\pi$/L $\times$ $\rho_x$ $\times$ ($l_z \cdot \bsigma^\pre_{\hat{y}})$) \\
		\tcp{The variables b$^\IJs_{xyz}$, c$^\IJs_{z}$ needs to be computed for every $(\is,\js)$ collision pair} 
		\ENDFOR
		\ENDFOR 
		\ENDFOR
		\ENDFOR
		\RETURN a,b$^\IJs$,c$^\IJs$
	\end{algorithmic}
	\caption{Pre-computation for Collision-Algorithm}
	\label{algo_fastSpectral_jaiswal2018_precompute}
\end{algorithm}

Next we outline the procedure for computing $\mathcal{Q}^\IJs$. Recall that our motive is to compute (\ref{eq_apndx_Qprojected})
\begin{itemize}
	\item First, we compute the forward Fourier transform of $\mathcal{F}^\Is_{i,\,l_1}$, and $\mathcal{F}^\Js_{i,\,l_2}$ to obtain $\hat{f}^\Is_l$ and $\hat{f}^\Js_m$ respectively. This is described in step 1 of Algo.~(\ref{algo_fastSpectral_jaiswal2018}). 
    \item Second, we compute $G^{\IJs+}(l,m)$ as per Eq.~(\ref{realGG}). Recall that $(\pi/L\;\rho\;l \cdot \bsigma)$ has been already precomputed and stored as $\mathrm{a}_{xyz}$. Also recall that $F(l+m,\rho,\bsigma)$ has been precomputed and stored as $\mathrm{b}^\IJs_{xyz}$. These can be reused to compute $G(l,m)$. This is described in step 2--8 of Algo.~(\ref{algo_fastSpectral_jaiswal2018}). In our implementation, we explicitly unroll the nested loops using Mako \cite{mako} templating engine, such that variables $\mathtt{t1,t2}$ in steps 4 and 5 are computed in a single kernel call (thereby requiring a space of $M N_\rho N^3$ each), and the FFT transforms in the step 6 are rather $M N_\q$ batched FFT transforms, each of size $N^3$.
    \item Third, in order to perform convolution for the loss-term $G^{\IJs-}(l,m)$, we prepare the variable \texttt{QG} in step 7 of Algo.~(\ref{algo_fastSpectral_jaiswal2018}). 
    \item Fourth, we perform convolutions to compute $\hat{\mathcal{Q}}^\IJs_k$ as in Eq.~(\ref{eq_apndx_Qprojected}). Recall that $G^{\IJs-}(m)$ has now been precomputed and stored as \texttt{QG}, and can be reused here. An inverse Fourier transform is then performed to obtain final $\mathcal{Q}^\IJs$. This is described in step 10 of Algo.~(\ref{algo_fastSpectral_jaiswal2018}). 
\end{itemize}

\begin{algorithm}[!ht]
\begin{algorithmic}[1]
	\renewcommand{\algorithmicrequire}{\textbf{Input:}}
	\renewcommand{\algorithmicensure}{\textbf{Output:}}
	\REQUIRE Number of points in each-direction of velocity mesh $N$, Distribution-functions $\mathcal{F}^\Is_{i,\,l_1}$ and $\mathcal{F}^\Js_{i,\,l_2}$ (size: $N^3$), number of points on half-sphere $M$, spherical quadrature weight $w_\bsigma$, Gauss quadrature-weights $w_\rho$ (size: $N_\q$), precomputed variable a (size: $MN_\q \times N^3$), precomputed variable b$^\IJs$ (size: $MN_\q \times N^3$), precomputed variable c$^\IJs$ (size: $N^3$), the kernel prefactor $\beta^\IJs$, normalized mass $\mi,\mj$ of species-pair $(\is,\js)$
	\ENSURE  Q \\
	\textit{Declare}: \\
		\quad \{t1,$\dots$,t3\} (each size: $N^3$); Q, QG (each size: $N^3$)
	\STATE Compute forward FFT: \\
		\quad FTf $\gets$ \fft($\mathcal{F}^\Is_{i,\,l_1}$) \\
		\quad FTg $\gets$ \fft($\mathcal{F}^\Js_{i,\,l_2}$) \\
	\tcp{Subscript x,y on symbols denote array-index}
    \tcp{Inner-most loop $r\in \{1,\dots,N^3\}$ has been ignored}
	\FOR {$x = 1$ to $N_\q$}
		\FOR {$y = 1$ to $M$} 
			\STATE t1 $\gets$ exp(\texttt{1i} $\times\;\mj/(\mi+\mj)\;\times$ a$_{xy}$) $\times$ FTf \\
			\tcp{Note: These are array-operations over $N^3$ ($z$ index)}
			\tcp{\texttt{1i} denotes the complex number $\sqrt{-1}$}
			\STATE t2 $\gets$ exp(\texttt{-1i} $\times\;\mi/(\mj+\mi)\;\times$ a$_{xy}$) $\times$ FTg \\
			\tcp{ifft denotes inverse FFT}
			\STATE t3 $\gets$ \fft(\ifft(t1)$\,\times\,$\ifft(t2)) \\
			\STATE QG $\gets$ QG + $({w_{\rho}})_x$ $\times$ $w_\bsigma$ $\times$ b$^\IJs_{xy}$ $\times$ t3 \\
		\ENDFOR
	\ENDFOR \\
	\tcp{real returns real part of complex number}
	\STATE Q = $\beta^\IJs\,\times\,$\texttt{real}( \ifft(QGs) - $\mathcal{F}^\Is_{i,\,l_1}\,\times\,$\ifft(c$^\IJs$ $\times$ FTg) )
	\RETURN Q
\end{algorithmic}
\caption{Collision-Algorithm Pseudo-code}
\label{algo_fastSpectral_jaiswal2018}
\end{algorithm}

\section{Micro-Benchmarks}
\label{section_benchmarks}
Verification for standard rarefied gas flows can be found in \cite{jaiswal2019dgfsMulti}. In the present work, we focus on the evaluation of the algorithmic behavior.

\subsection{Hardware Configuration}
Serial and parallel implementations of multi-species DGFS solver are run on 15-node Brown-GPU RCAC cluster at Purdue University. Each node is equipped with two 12-core Intel Xeon Gold 6126 CPU, and three Tesla-P100 GPU. The operating system used is 64-bit  CentOS 7.4.1708 (Core) with NVIDIA Tesla-P100 GPU accompanying CUDA driver 8.0 and CUDA runtime 8.0. The GPU has 10752 CUDA cores, 16GB device memory, and compute capability of 6.0. The solver has been written in Python/PyCUDA and is compiled using OpenMPI 2.1.0, g++ 5.2.0, and  nvcc 8.0.61 compiler with third level optimization flag. All the simulations are done with double precision floating point values. %

\subsection{Spatially homogeneous case: Krook-Wu exact solution}
For constant collision kernel, an exact solution to the spatially homogeneous multi-species Boltzmann equation can be constructed (see \cite{krook1977exact}). We use this solution to verify the accuracy of the proposed fast spectral method for approximating the collision operator. Considering a binary mixture ($n=2$; $\is=1,2$), the equation simplifies to
\begin{align}
\partial_t f^\Is = \sum_{\js=1}^{2}\int_{\mathbb{R}^3}\int_{S^{2}} B_{\ijs}\left[f^{\Is}(v')f^{\Js}(v_*')-f^{\Is}(v)f^{\Js}(v_*)\right]\rd{\bsigma}\,\rd{v_*},
\label{eq_bkw_pde}
\end{align}
where $B_{\ijs}=B_{\jis}:=\frac{\lambda_{\jis}}{4\pi n^{\Js}}$ and $\lambda_{\ijs}$ is some positive constant. The exact solution is given by
\begin{align}
f^\Is(t,v) = n^\Is \Bigg( \frac{m_\is}{2\pi K} \Bigg)^{3/2} \exp{\Bigg(-\frac{m_\is v^2}{2K}\Bigg)} \Bigg((1-3Q_\is) + \frac{m_\is}{K} Q_\is v^2 \Bigg),
\label{eq_bkw_f}
\end{align}
where
\begin{align}
& \mu=\frac{4m_1 m_2}{(m_1+m_2)^2}, \quad \tau_1 = \lambda_{22} - \lambda_{21} \mu (3-2\mu), \quad \tau_2 = \lambda_{11} - \lambda_{12} \mu (3-2\mu), \nonumber \\
& A = \frac{1}{6} \Bigg( \lambda_{11} + \lambda_{21} \mu\left(3 -2 \mu \frac{\tau_2}{\tau_1}\right) \Bigg), \quad B = \frac{1}{3} \Bigg( \lambda_{11} \tau_1 + \lambda_{21} \mu (3 -2 \mu) \tau_2 \Bigg),\nonumber \\
& Q(t) = \frac{A}{A\exp(A t)-B}, \quad Q_\is(t) = \tau_\is Q(t),   \nonumber\\
&K(t) = \frac{n^\one + n^\two}{ (n^\one + n^\two) + 2( n^\one \tau_1 + n^\two \tau_2) Q(t) }.
\end{align}
Furthermore, the following condition needs to be satisfied
\begin{equation}
(\tau_1-\tau_2)\left(2\mu^2\left(\frac{\lambda_{21}}{\tau_1}-\frac{\lambda_{12}}{\tau_2}\right)-1\right)=0.
\end{equation}
For simplicity, we choose $n^\one=n^\two=1$, $\lambda_{11}=\lambda_{22}=1$, $\lambda_{12}=\lambda_{21}=1/2$ but vary the mass ratio $m_1/m_2$ in the following tests.

It is also helpful to take the derivative of eqn.~(\ref{eq_bkw_f}), which yields
\begin{align}
& \partial_t f^\Is = f^\Is \Bigg( -\frac{3}{2K}K' + \frac{m_\is\;v^2}{2 K^2} K'\Bigg) \nonumber \\
&+ n^\Is \Bigg(\frac{m_\is}{2\pi K}\Bigg)^{3/2} \exp{\Bigg(-\frac{m_\is v^2}{2K}\Bigg)} \Bigg(-3Q'_\is + \frac{m_\is}{K} Q'_\is v^2 - \frac{m_\is}{K^2} K' Q_\is v^2 \Bigg) \nonumber\\
&:=\sum_{\js=1}^2 \mathcal{Q}^{\IJs}(f^{\is},f^{\js}),
\end{align}
where
\begin{align}
&Q'(t)=-\frac{A^3\exp(At)}{(A\exp(At)-B)^2}, \quad Q_\is'(t)=\tau_\is Q'(t), \quad \nonumber \\ &K'(t)=-\frac{2(n^{(1)}+n^{(2)})(n^{(1)}\tau_1+n^{(2)}\tau_2)}{[(n^{(1)}+n^{(2)})+2(n^{(1)}\tau_1+n^{(2)}\tau_2)Q(t)]^2}Q'(t).
\end{align}
This allows us to check the accuracy of the collision solver without introducing time discretization error.

\begin{table*}[!ht]
\centering
\setlength{\tabcolsep}{0.3em}
\begin{tabular}{@{}cc|ccc|ccc|ccc|ccc@{}}
\toprule
\multirow{3}{*}{$N$} & \multirow{3}{*}{$N_\rho$} & \multicolumn{6}{c|}{$m_\js/m_\is=1$} & \multicolumn{6}{c}{$m_\js/m_\is=4$}  \\
   &   & \multicolumn{3}{c|}{$M=6$} & \multicolumn{3}{c|}{$M=12$} & \multicolumn{3}{c|}{$M=6$} & \multicolumn{3}{c}{$M=12$}  \\
   &   &     $time\,(s)$ & $\mathcal{E}^\one$ & $\mathcal{E}^\two$ & $time\,(s)$ & $\mathcal{E}^\one$ & $\mathcal{E}^\two$ & $time\,(s)$ & $\mathcal{E}^\one$ & $\mathcal{E}^\two$ & $time\,(s)$ & $\mathcal{E}^\one$ & $\mathcal{E}^\two$ \\ \midrule
16 & 4 & 0.00039 & 3.27e-03 & 3.27e-03 & 0.00050 & 1.77e-03 & 1.77e-03 & 0.00039 & 4.80e-03 & 1.22e-03 & 0.00050 & 3.63e-03 & 2.47e-04 \\
 & 8 & 0.00056 & 3.73e-03 & 3.73e-03 & 0.00085 & 2.00e-03 & 2.00e-03 & 0.00050 & 4.96e-03 & 1.33e-03 & 0.00093 & 3.62e-03 & 2.42e-04 \\
 & 16 & 0.00084 & 3.73e-03 & 3.73e-03 & 0.00147 & 2.00e-03 & 2.00e-03 & 0.00084 & 4.96e-03 & 1.33e-03 & 0.00148 & 3.62e-03 & 2.42e-04 \\
24 & 6 & 0.00068 & 1.37e-04 & 1.37e-04 & 0.00114 & 1.01e-04 & 1.01e-04 & 0.00068 & 1.81e-03 & 2.06e-02 & 0.00114 & 1.79e-03 & 5.15e-03 \\
 & 12 & 0.00117 & 1.49e-04 & 1.49e-04 & 0.00209 & 9.64e-05 & 9.64e-05 & 0.00114 & 2.12e-03 & 1.87e-02 & 0.00210 & 2.13e-03 & 6.01e-03 \\
 & 24 & 0.00210 & 1.49e-04 & 1.49e-04 & 0.00401 & 9.64e-05 & 9.64e-05 & 0.00210 & 2.12e-03 & 1.87e-02 & 0.00401 & 2.13e-03 & 6.01e-03 \\
32 & 8 & 0.00159 & 3.04e-05 & 3.04e-05 & 0.00287 & 2.51e-05 & 2.51e-05 & 0.00157 & 1.54e-04 & 1.62e-02 & 0.00286 & 1.52e-04 & 1.13e-02 \\
 & 16 & 0.00286 & 3.17e-05 & 3.17e-05 & 0.00541 & 2.45e-05 & 2.45e-05 & 0.00286 & 5.91e-05 & 1.69e-02 & 0.00542 & 5.87e-05 & 1.03e-02 \\
 & 32 & 0.00543 & 3.17e-05 & 3.17e-05 & 0.01057 & 2.45e-05 & 2.45e-05 & 0.00542 & 5.91e-05 & 1.69e-02 & 0.01059 & 5.87e-05 & 1.03e-02 \\
40 & 10 & 0.00328 & 1.38e-06 & 1.38e-06 & 0.00626 & 1.26e-06 & 1.26e-06 & 0.00326 & 5.53e-05 & 4.31e-03 & 0.00626 & 5.56e-05 & 4.35e-03 \\
 & 20 & 0.00625 & 9.35e-07 & 9.35e-07 & 0.01226 & 8.10e-07 & 8.10e-07 & 0.00626 & 5.01e-05 & 4.29e-03 & 0.01222 & 4.97e-05 & 4.54e-03 \\
 & 40 & 0.01227 & 9.35e-07 & 9.35e-07 & 0.02446 & 8.10e-07 & 8.10e-07 & 0.01219 & 5.01e-05 & 4.29e-03 & 0.02431 & 4.97e-05 & 4.54e-03 \\
48 & 12 & 0.00656 & 1.04e-07 & 1.04e-07 & 0.01289 & 9.99e-08 & 9.99e-08 & 0.00658 & 8.46e-06 & 5.76e-04 & 0.01291 & 8.45e-06 & 5.93e-04 \\
 & 24 & 0.01290 & 1.05e-07 & 1.05e-07 & 0.02556 & 9.95e-08 & 9.95e-08 & 0.01291 & 7.17e-06 & 5.80e-04 & 0.02561 & 7.51e-06 & 6.09e-04 \\
 & 48 & 0.02545 & 1.05e-07 & 1.05e-07 & 0.05169 & 9.95e-08 & 9.95e-08 & 0.02550 & 7.17e-06 & 5.80e-04 & 0.05215 & 7.51e-06 & 6.09e-04 \\
56 & 14 & 0.01204 & 9.80e-08 & 9.80e-08 & 0.02350 & 9.79e-08 & 9.79e-08 & 0.01202 & 5.22e-06 & 2.32e-04 & 0.02352 & 4.08e-06 & 1.88e-04 \\
 & 28 & 0.02354 & 9.80e-08 & 9.80e-08 & 0.04667 & 9.79e-08 & 9.79e-08 & 0.02353 & 5.09e-06 & 2.24e-04 & 0.04662 & 3.97e-06 & 1.87e-04 \\
 & 56 & 0.04664 & 9.80e-08 & 9.80e-08 & 0.09303 & 9.79e-08 & 9.79e-08 & 0.04674 & 5.09e-06 & 2.24e-04 & 0.09313 & 3.97e-06 & 1.87e-04 \\
\bottomrule
\end{tabular}
\caption{Efficiency and accuracy $L^\infty$ error $\mathcal{E}^\Is = \| \partial_t f_{analytical}^\Is - \partial_t f_{numerical}^\Is \|,\;\is=\{1,2\}$ for spatially homogeneous Krook-Wu solution at $t=5.5$ for different mass-ratios. $N$, $N_\rho$, and $M$ respectively, denote the number of discretization points in the velocity space, number of Gauss quadrature points in the radial direction, and number of discretization points on full sphere. A fixed velocity domain $[-12,\;12]^3$ has been used for all the mass-ratios.}
\label{tab_bkw_error}
\end{table*}

Table~\ref{tab_bkw_error} shows the $L^\infty$ norm between the numerical and analytical $\partial f^\Is/\partial t$. For different mass ratios, we have considered the cases with $N= \{16,\,24,\,32,\,40,\,48,\,56\}$ points in each velocity dimension; and $M=6$, $12$ spherical design quadrature points on the full sphere. A good agreement between analytical and numerical solutions is evident from the table. At a fixed $N$, with increase in mass ratio, the error norm increases. In particular, increase in $M$ does not considerably affect the solution due to the isotropic nature of the distribution function. Note that, in the fast spectral decomposition, since the integral oscillates roughly on $O(N)$, the total number of Gauss–Legendre quadrature points $N_\rho$ in the radial direction should be on order of $O(N)$. As per \cite{GHHH17}, a more precise estimate is $\approx 0.8 N$. However, there is no good rule to select optimal $N_\rho$. We observe that the error is relatively unaffected upon reducing $N_\rho$ from $N$ to $N/2$. However, we note that $N_\rho=N$ is a safer choice.

From a computational viewpoint, the simulation time is independent of the mass ratio. On increasing the number of discretization points on the sphere $M$, the computational cost approximately doubles--however, we do observe the effect of loop unrolling for smaller $N$. Likewise, the computational cost approximately doubles on increasing the number of quadrature points $N_\rho$. This establishes that the algorithm is linear in both $M$ and $N_\rho$. 

\subsection{Spatially in-homogeneous case: Couette flow}
The aforementioned methodology allows us to compute the collision operator efficiently. To solve the fully spatial in-homogeneous equation (\ref{eqn_bzeNonDim}), we also need an accurate and efficient spatial and time discretization. Here, we adopt the the Runge-Kutta discontinuous Galerkin (RKDG) approach--widely used for hyperbolic systems--as adapted in \cite{JAH19,jaiswal2019dgfsMulti} for Boltzmann equation. The details of the discretization can be found in \cite{JAH19,jaiswal2019dgfsMulti}. We mention that evaluation of collision operator consumes $> 98\%$ of computation time, and hence, in the present work, we focus on the collision operator behavior. More details on spatial-temporal RKDG discretization on GPU can be found in \cite{klockner2009nodal,witherden2014pyfr}. We restrict our discussion and benchmarks to 1-D flow problems for brevity\footnote{Discussion and benchmarks for higher 2D/3D spatial dimension shall be presented in the extended version of this manuscript.}.

\subsubsection{Verification}
For general Boltzmann equation (\ref{eqn_bzeNonDim}), analytical solutions do not exist. Therefore, we compare our results with widely accepted direct simulation Monte Carlo (DSMC) \cite{Bird} method. We want to emphasize that DSMC is a stochastic method for solution of the N-particle master kinetic equation which converges to the Boltzmann equation in the limit of infinite number of particles \cite{wagner1992convergence}. 

\begin{figure}[!ht]
	\centering
    \begin{tikzpicture}
       \def\pr{1.23};
       \foreach \x in {1,...,5}
         \fill ({-\pr + (\pr)^\x},0cm) circle (0.05cm);

       \foreach \x in {1,...,5}
         \fill ({ 4 + \pr - (\pr)^(\x)},0cm) circle (0.05cm);

       \draw(0,0) -- (4,0) ;
       \draw[blue] (0.0,0.0) circle (0.1cm);
       \draw[red] (4.0,0.0) circle (0.1cm);

       \draw[-latex] (0.0,-0.5) -- (0.0, 0.0) node[below, yshift=-0.5cm] {$\mathbf{u}_l,\;T_l$};
       \draw[-latex] (4.0,-0.5) -- (4.0, 0.0) node[below, yshift=-0.5cm] {$\mathbf{u}_r,\;T_r$};

       \draw[-latex] (-2.0,-1) -- (-1.0, -1) node[anchor=north] {x};
       \draw[-latex] (-2.0,-1) -- (-2.0, 0) node[anchor=south] {y};
    \end{tikzpicture}
	\caption{Numerical setup for 1D Couette flow.}
	\label{fig_couetteFlowSchematic}
\end{figure}
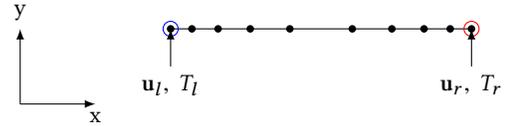

\begin{figure}[!ht]
\begin{subfigure}[b]{.5\textwidth}
  \centering
  \includegraphics[width=80mm]{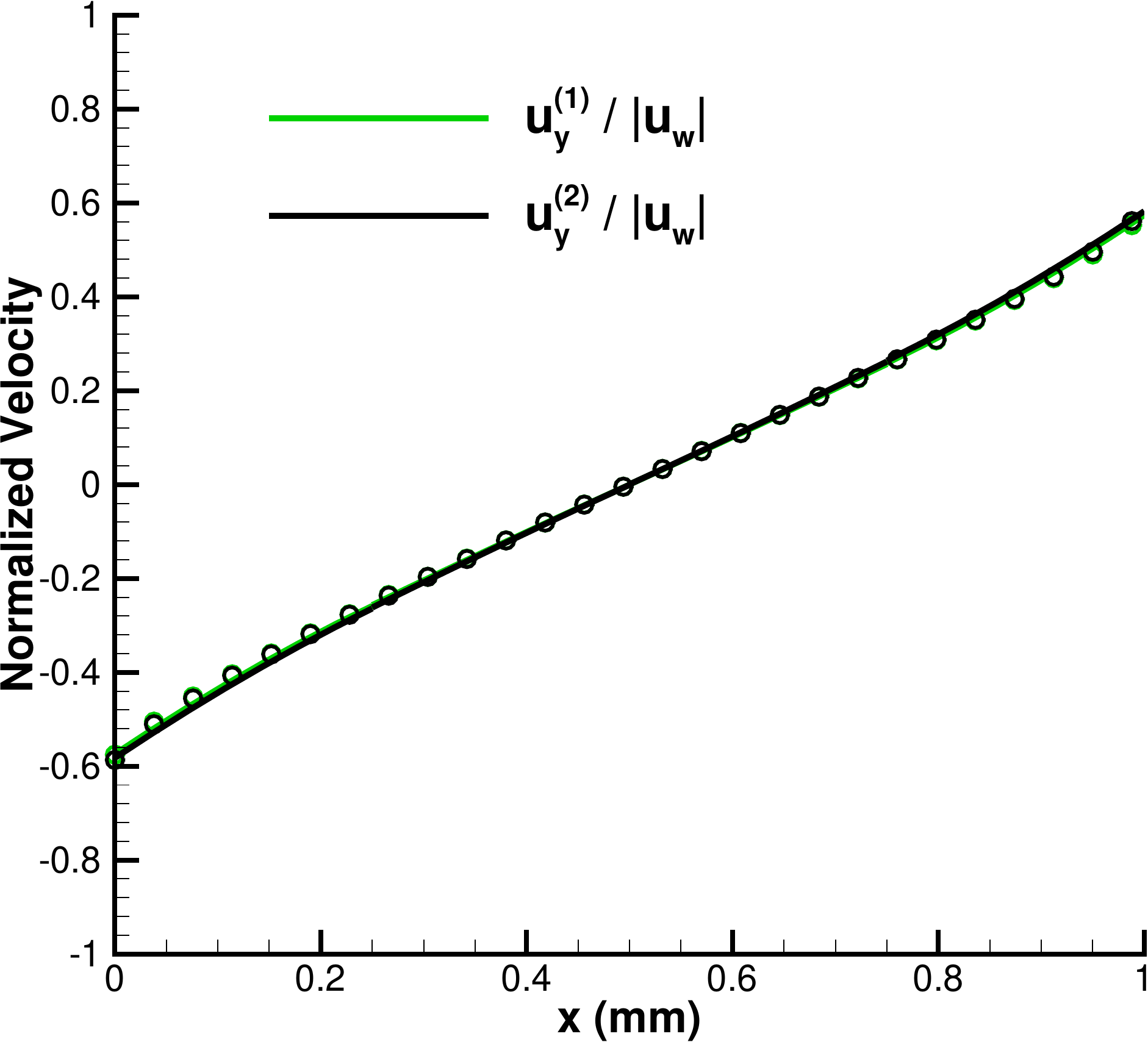}
  \caption{normalized $y$-component of velocity}
  \label{fig_couetteVSS_U}
\end{subfigure}%

\begin{subfigure}[b]{.5\textwidth}
  \centering
  \includegraphics[width=80mm]{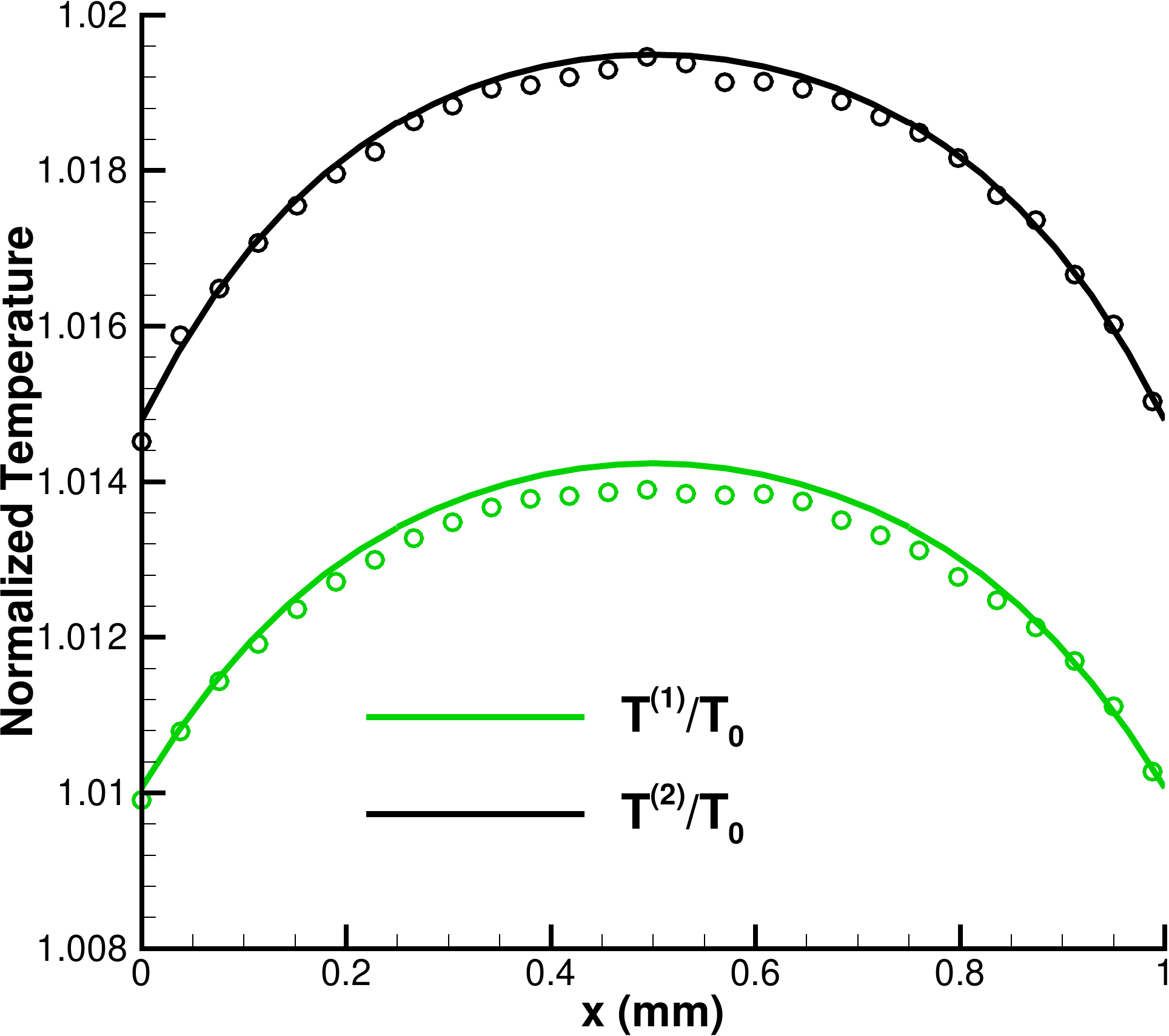}
  \caption{normalized temperature}
  \label{fig_couetteVSS_T}
\end{subfigure}
\caption{Variation of normalized $y$-velocity, and temperature along the domain for Couette flow (Case C-01) obtained with DSMC and DGFS using VSS collision kernel for Argon-Krypton mixture. Symbols denote DSMC solutions, and lines denote DGFS solutions.}
\label{fig_couetteVSS_U_T}
\end{figure}

In the current test case, we consider the effect of velocity gradient on the solution. The coordinates are chosen such that the walls are parallel to the $y$ direction and $x$ is the direction perpendicular to the walls. The geometry as well as boundary conditions are shown in Figure \ref{fig_couetteFlowSchematic}. Figure~\ref{fig_couetteVSS_U_T} illustrates the velocity and temperature along the domain length for both species, wherein we observe an excellent agreement between DGFS and DSMC. The small discrepancies, however, are primarily due to: a) statistical fluctuations inherent to the Monte Carlo methods, b) practical limitations on number of particles used in DSMC simulations. From a computational viewpoint, the present DGFS simulations on a single GPU took 138 seconds to acquire the steady state, in contrast to 26086.45 sec on 24 processors for DSMC simulations as reported in \cite{jaiswal2019dgfsMulti}, for achieving comparable accuracy.

\begin{table}[!ht]
\centering
\begin{tabular}{@{}lcccc@{}}
\toprule
Parameter & Case C-01 \\ 
\midrule
Molecular mass: $\{m_1,\,m_2\}$ ($\times 10^{27}\,kg$) & $\{66.3$,\,$139.1\}$ \\ 
Non-dim physical space & $[0,\,1]$ \\ 
Non-dim velocity space & $[-7,\,7]^3$ \\ 
$\{N^3,\,N_\rho,\,M\}$ & $\{32^3,\,8,\,12\}$ \\
Spatial elements & $4$ \\
DG order & 3 \\
Time stepping & Euler \\
Viscosity index: $\omega_{\{11,\,12,\,21,\,22\}}$ & $\{0.81,\,0.805,\,0.805,\,0.8\}$ \\
Scattering parameter: $\alpha_{\{11,\,12,\,21,\,22\}}$ & $\{1.4,\,1.36,\,1.36,\,1.32\}$ \\
Ref. diameter: $d_{\text{ref},\ijs}$ ($\times 10^{10} m$) & $\{4.11,\,4.405,\,4.405,\,4.7\}$ \\
Ref. temperature: $T_{\text{ref},\ijs}$ ($K$) & $\{273\}$ \\
Characteristic mass: $m_0$ ($\times 10^{27}\,kg$) & $66.3$ \\
Characteristic length: $H_0$ ($mm$) & 1 \\
Characteristic velocity: $\mathbf{u}_0$ ($m/s$) & 337.2 \\
Characteristic temperature: $T_0$ ($K$) & 273 \\
Characteristic number density: $n_0$ ($m^{-3}$) & $1.680 \times 10^{21}$ \\
\midrule
\multicolumn{2}{l}{Initial conditions} \\
Velocity: $\mathbf{u}$ ($m/s$) & 0 \\
Temperature: $T$ ($K$) & 273 \\
Number density: $n^\one$ ($m^{-3}$) & $1.680\times10^{21}$ \\
Number density: $n^\two$ ($m^{-3}$) & $8.009\times10^{20}$ \\
Knudsen number: $(\Kn_{11},\,\Kn_{22})$ & $(0.793,\,0.606)$  \\
Knudsen number: $(\Kn_{12},\,\Kn_{21})$ & $(0.803,\,0.555)$ \\
\midrule
\multicolumn{2}{l}{Left wall (purely diffuse) boundary conditions (subscript $l$)} \\
Velocity: $\mathbf{u}_l$ ($m/s$) & $(0,\,-50,\,0)$ \\
Temperature: $T_l$ ($K$) & 273 \\
\midrule
\multicolumn{2}{l}{Right wall (purely diffuse) boundary conditions (subscript $r$)} \\
Velocity: $\mathbf{u}_r$ ($m/s$) & $(0,\,+50,\,0)$ \\
Temperature: $T_r$ ($K$) & 273 \\
\bottomrule
\end{tabular}
\caption{Numerical parameters for Couette flow \cite{jaiswal2019dgfsMulti}. Based upon our observations from Table~\ref{tab_bkw_error}, we have used $N_\rho=8$, in contrast to $N_\rho=32$ used in \cite{jaiswal2019dgfsMulti}. This does not affect the recovered bulk properties as illustrated in Fig.~\ref{fig_couetteFlowSchematic}, however, it speeds up the computation by a factor of 4.}
\label{tab_couette_conditions}
\end{table}

\subsubsection{Scaling Behavior}
\begin{table*}
 	\caption{\normalfont Performance of the solver for Couette flow test cases. The phase-space is defined using a convenient triplet notation \textbf{N$_e$/K/N$^3$}, which corresponds to \textbf{N}$_e$ elements in physical space, \textbf{K} order DG (equivalently $N_p=K-1$ order polynomial for 1-D domain), and \textbf{N}$^3$ points in velocity space. $n$\textbf{G} ($n>1$) denotes GPU/CUDA/MPI/parallel execution on $n$ GPUs shared equally across $(n/3)$ nodes. Work units represent the total simulation time for first 52 timesteps. Efficiency is defined as ratio ($1$\textbf{G}/$n$\textbf{G})/n, where 1\textbf{G} and $n$\textbf{G} are execution-times on one GPU and $n$ GPU respectively. $M=12$ and $N_\rho=8$ is used for all cases.}
\centering  
\begin{tabular}{@{}lrrrrrrrrrrrrr@{}}
\toprule
\multirow{2}{*}{Phase space} & \multicolumn{7}{c}{Work Units (s)} & \multicolumn{6}{c}{Efficiency} \\  \cmidrule(l){2-8} \cmidrule(l){9-14} 
    & 1G & 3G & 6G & 9G & 12G & 24G & 36G & 1G/3G & 1G/6G & 1G/9G & 1G/12G & 1G/24G & 1G/36G
    \\ \midrule
$72/3/20^3$ & 47.580 & 16.155 & 8.339 & 5.698 & 4.392 & 2.423 & 1.774 & 0.98 & 0.95 & 0.93 & 0.90 & 0.82 & 0.84\\
$72/3/32^3$ & 126.601 & 42.616 & 21.551 & 14.563 & 11.038 & 5.784 & 4.030 & 0.99 & 0.98 & 0.97 & 0.96 & 0.91 & 0.98\\
$72/3/48^3$ & 391.943 & 131.081 & 65.913 & 44.218 & 33.513 & 17.224 & 11.621 & 1.00 & 0.99 & 0.98 & 0.97 & 0.95 & 1.05 \\
$72/6/20^3$ & 94.682 & 31.957 & 16.197 & 10.944 & 8.331 & 4.392 & 3.079 & 0.99 & 0.97 & 0.96 & 0.95 & 0.90 & 0.96 \\
$72/6/32^3$ & 253.016 & 84.834 & 42.741 & 28.697 & 21.703 & 11.158 & 7.693 & 0.99 & 0.99 & 0.98 & 0.97 & 0.94 & 1.03 \\
$72/6/48^3$ & 782.343 & 261.601 & 131.217 & 87.755 & 66.009 & 33.520 & 22.509 & 1.00 & 0.99 & 0.99 & 0.99 & 0.97 & 1.09 \\
\midrule
$216/3/20^3$ & 141.754 & 47.641 & 24.033 & 16.182 & 12.326 & 6.356 & 4.388 & 0.99 & 0.98 & 0.97 & 0.96 & 0.93 & 1.01 \\
$216/3/32^3$ & 378.956 & 126.853 & 63.676 & 42.636 & 32.066 & 16.295 & 11.041 & 1.00 & 0.99 & 0.99 & 0.98 & 0.97 & 1.07 \\
$216/3/48^3$ & 1172.907 & 391.916 & 196.439 & 131.153 & 98.538 & 49.652 & 33.471 & 1.00 & 1.00 & 0.99 & 0.99 & 0.98 & 1.10 \\
$216/6/20^3$ & 283.091 & 94.737 & 47.679 & 31.903 & 24.060 & 12.262 & 8.320 & 1.00 & 0.99 & 0.99 & 0.98 & 0.96 & 1.06 \\
$216/6/32^3$ & 759.149 & 253.498 & 127.004 & 84.932 & 63.780 & 32.212 & 21.672 & 1.00 & 1.00 & 0.99 & 0.99 & 0.98 & 1.09 \\
$216/6/48^3$ & 2347.099 & 783.642 & 392.470 & 261.817 & 196.552 & 98.680 & 66.018 & 1.00 & 1.00 & 1.00 & 1.00 & 0.99 & 1.11 \\
\bottomrule
\end{tabular}
\label{tab_ns_fc_speedup}
\end{table*}

The simulations are carried out for different test-cases by varying element-count ($N_e$), polynomial approximation order ($N_p=K-1$), and velocity-space sizes ($N$). The spatial elements are distributed to $p$ processors using the well-known linear domain-decomposition strategy requiring sharing of $O(p N^3)$ floating-point during MPI communication phase. Speed up obtained with multi-GPU solver is presented in Table~(\ref{tab_ns_fc_speedup}). As evident from the table, the acceleration due to GPU parallelization increases with increase in the size of computational grid. More specifically, the increase in $N_e$ and $K$ have small-effect on overall speedup which suggests that DG-operators (for instance derivative, time-evolution) are rather computationally inexpensive operations. On the other hand, increase in velocity-grid improves the observed speedup. The weak/strong scaling behavior is also evident from the table.

\subsubsection{Flat profile}
\begin{figure*}[!ht]
\begin{subfigure}{.5\textwidth}
  \centering
  \includegraphics[width=80mm]{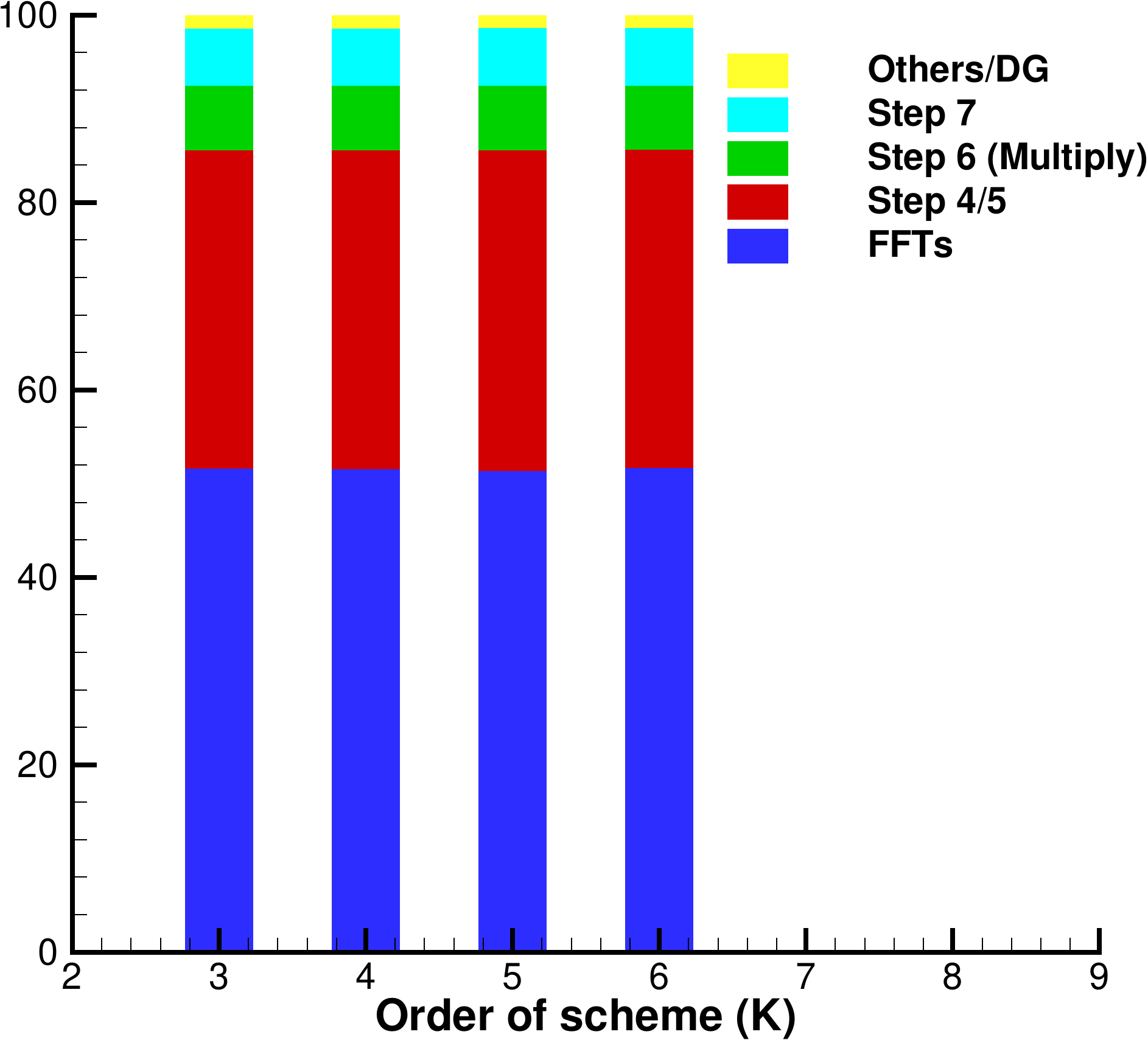}
  \caption{$N_e=72$, $N^3=20^3$}
  \label{fig_couetteVSS_profile_72k20}
\end{subfigure}%
\begin{subfigure}{.5\textwidth}
  \centering
  \includegraphics[width=80mm]{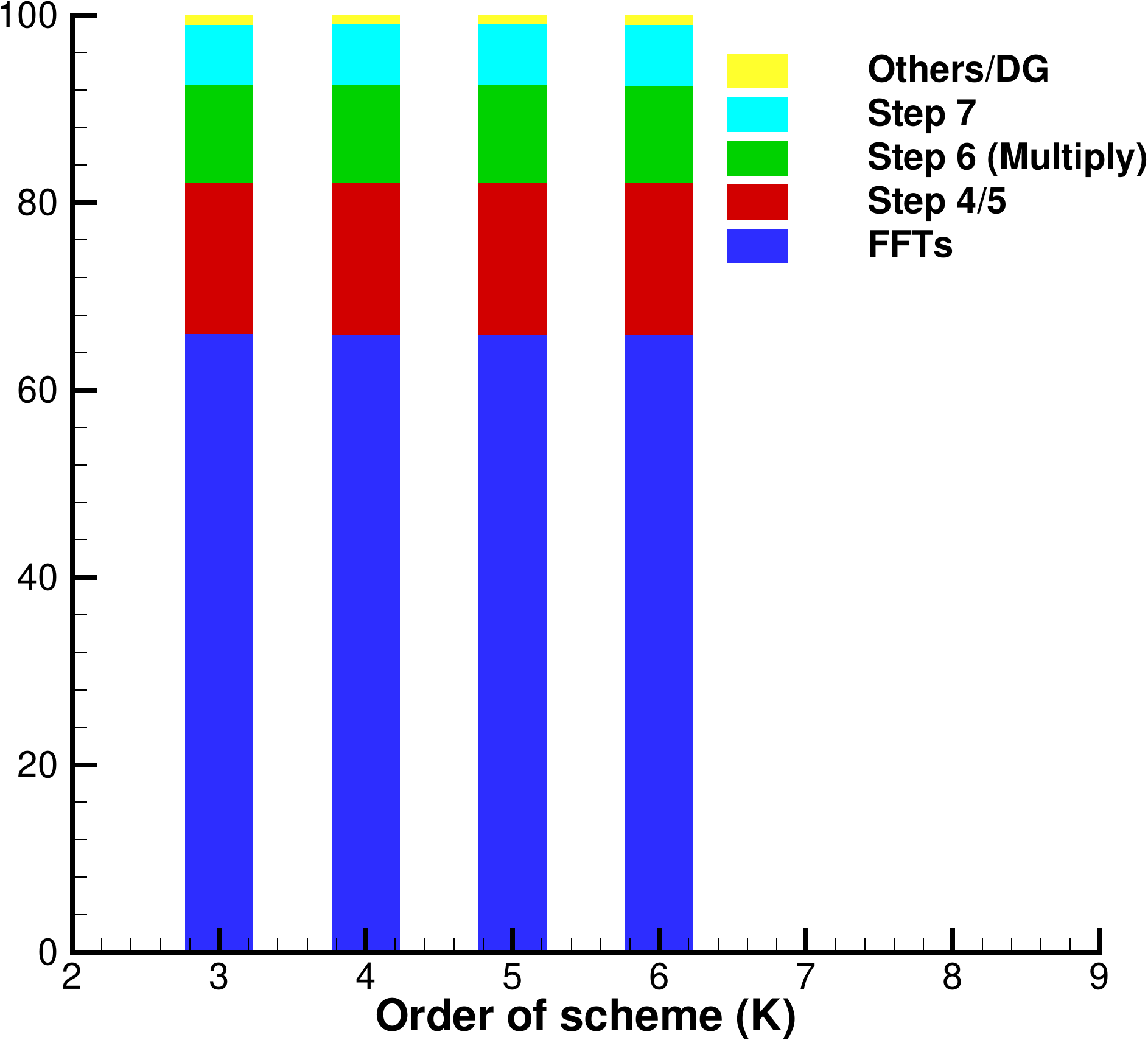}
  \caption{$N_e=72$, $N^3=32^3$}
  \label{fig_couetteVSS_profile_72k32}
\end{subfigure}
\caption{Percentage of time spent in various parts of Algo.~\ref{algo_fastSpectral_jaiswal2018} vs. order of DG scheme (K). For both $N^3=20^3$ and $N^3=32^3$, the collision operator consumes $>98\%$ of the simulation time. }
\label{fig_couetteVSS_profile}
\end{figure*}

Recall that the fast Fourier spectral collision operator algorithm~\ref{algo_fastSpectral_jaiswal2018} is split into multiple parts. It is therefore interesting to see what performance level is attained by each part of the operator.  Fig~(\ref{fig_couetteVSS_profile}) presents the percentage of time spent in various parts of Algo.~\ref{algo_fastSpectral_jaiswal2018} vs. order of DG scheme (K). First, we note that the DG operators denoted in yellow, requires $~1\%$ of the total simulation time. The collision operator, however, consumes nearly $>98\%$ of the total time for both $N^3=20^3$ and $N^3=32^3$. 

\section{Conclusions}
\label{section_conclusions}
We have presented an implementation of the multi-species Discontinuous Galerkin Fast Spectral (DGFS) method for solution of \textit{multi-species} monoatomic full Boltzmann equation on multi-GPU/multi-CPU architectures. The DG-type formulation employed in the present work has advantage of having high-order accuracy at the element-level, and its element-local compact nature (and that of our collision algorithm) enables effective parallelization on massively parallel architectures. For verification and benchmarks, we carry out simulations for spatially homogeneous BKW, and Couette flow problems. Parallel efficiency close to 0.95 is observed on a 36 GPU multi-node/multi-GPU system. An important key observation is that the efficiency can be maintained provided we have enough work on each processor. It is this speedup that now allows researchers to solve problems within a day that would otherwise take months on traditional CPUs. Future work directions include, assessment of the implementation beyond thousand cores. Extending the implementation to general 2D/3D mixed grids coupled with adaptivity in physical and velocity spaces, is an interesting direction as well. 

\bibliographystyle{ACM-Reference-Format}
\bibliography{pasc-sigconf}

\end{document}